\documentclass[10pt,preprint,showpacs,preprintnumbers,twocolumn,
               superscriptaddress,amsmath,amssymb,floatfix]{revtex4}
\usepackage{graphicx,epsfig,latexsym,amssymb}
\usepackage{multirow,amsmath,array,booktabs}
\usepackage[section]{placeins}
\usepackage{bm}
\usepackage{xcolor}

\usepackage{natbib}

\renewcommand{\emph}[1]{\textit{\textbf{#1}}}

\begin{document}

\title{Informal Science Education and Career Advancement}

\author{Michael S. Smith}
\email{smithms@ornl.gov}
\affiliation{Physics Division, Oak Ridge National Laboratory, Oak Ridge, TN, 37831-6354, USA}

\author{Claudia Fracchiolla}
\email{fracchiolla@aps.org}
\affiliation{American Physical Society, 1 Physics Ellipse, College Park, MD, 20740-3844, USA}

\author{Sean Fleming}
\affiliation{College of Earth, Ocean, and Atmospheric Sciences, Oregon State university, Corvallis, OR, 97331, USA; Department of Earth, Ocean, and Atmospheric Sciences, University of British Columbia, Vancouver, BC, Canada; and National Water and Climate Center, Natural Resources Conservation Service, US Department of Agriculture, Portland, OR, 97232, USA}

\author{Arturo Dominguez}
\affiliation{Princeton Plasma Physics Laboratory, 100 Stellarator Rd, Princeton, NJ, 08540, USA}

\author{Alexandra Lau}
\affiliation{American Physical Society, 1 Physics Ellipse, College Park, MD, 20740-3844, USA}

\author{Shannon Greco}
\affiliation{Princeton Plasma Physics Laboratory, 100 Stellarator Rd, Princeton, NJ, 08540, USA}

\author{Don Lincoln}
\affiliation{Fermi National Accelerator Laboratory, Batavia, Illinois, 60510, USA}

\author{Eleni Katifori}
\affiliation{Department of Physics \& Astronomy, University of Pennsylvania, 209 South 33rd Street, Philadelphia, PA, 19104-6396, USA}

\author{William Ratcliff}
\affiliation{NIST, 100 Bureau Dr., Gaithersburg, MD, 20889, USA and Dept. of Materials Science, Univ. of Maryland, College Park, MD, 20742, USA}

\author{Maria Longobardi}
\affiliation{University of Basel, Department of Physics, Basel, Switzerland}

\author{Maajida Murdock}
\affiliation{Dept of Physics \& Engineering Physics, Morgan State University, 1700 East Cold Spring Lane, Baltimore, MD, 21251, USA}

\author{Mustapha Ishak}
\affiliation{Department of Physics, The University of Texas at Dallas, Richardson, TX, 75080, USA}

\date{\today}

\begin{abstract}
This document supports a proposed APS statement that encourages academic, research, and other institutions to add the participation in informal science education activities to the criteria they use for hiring and career advancement decisions. There is a prevalent attitude that the time spent by a researcher on these activities takes time away from research efforts that are more valued by their peers and their institution. To change this mindset, we enumerate the many benefits of informal science education activities to the public, to researchers, to their institutions, and to the field of physics. We also suggest aspects of these activities that may be considered by institutions in evaluating informal educational efforts for recruitment and career advancement decisions.
\end{abstract}


\maketitle

\section{Introduction and Draft APS statement}

Informal science education activities have long played an important role in helping create and recruit the next generation of scientists, in communicating both exciting results and fundamental concepts to the world, in building public trust in science and scientists, in raising the profile of research and academic institutions, in justifying the use of taxpayer funds for research, and in securing new funds for important projects \cite{NAP12190}. Despite their value, the efforts of physicists to initiate and participate in informal education activities are often not considered in hiring decisions or in career advancement decisions by their institution, as evident from faculty handbooks ({\it e.g.}, \cite{HARV2021}. In order to recognize the benefits of informal educational efforts, and to motivate more physicists to participate in them in the future, the APS CIP has drafted a statement that advocates that engagement-based efforts be considered in recruitment and career advancement decisions by the facilitators' home institutions.

\vskip 0.1cm
\noindent The text of the draft statement is as follows:

\vskip 0.1cm

{\it Systematic, ongoing, respectful, lively, two-way conversations with the public -- that is, public engagement on science -- is critical to the field of physics, including the public image of institutions hosting physics research and education, the recruitment and diversity of new generations of physicists, the scientific interest and literacy of the general public and in turn their support of physics and science more generally, and the success of physics-based applied research and development undertaken in response to specific practical societal needs. Overall, however, such activities tend to receive inconsistent and in some cases negligible evaluation during hiring, performance assessments, and promotions of physicists. APS therefore strongly supports participation in informal education activities, and supports that such participation should be considered in recruiting and promotion decisions, including tenure decisions at universities and other forms of career advancement at non-academic institutions as appropriate.} 

\vskip 0.1cm

In the sections below, we will define terminology for informal science education efforts and describe the different types of engagement activities, and we identify the stakeholders in these efforts and the value to each. We also give examples of some successful programs, and we suggest aspects of these activities that may be considered by institutions in evaluating informal educational efforts for hiring and career advancement decisions.
 
\section{Description of Informal Education Terms and Activities}

To properly frame our discussion, we begin by defining and then characterizing informal science education. In contrast to classroom learning, informal education takes place outside of formal learning structures, often lacks a framework and defined objectives, and includes a broad range of activities \cite{NAP12190, ainsworth2010formal}. Informal learning happens continuously, as it is based on the individual's experience. In some cases like watching TV or listening to the radio, the learning may be happening without the individual explicitly choosing it. In other cases like visiting museums or zoos, reading books, attending summer camps, or going to a public lecture, it can be more deliberate. Informal science education is widely available and regularly consumed by the public, and in fact plays an important role in people's general understanding of science. Informal science is often characterized by ``free-choice'', learner-led activities, which are unbounded by standards or metrics, and which have an emphasis on the participant's curiosity and excitement \cite{falk2005using, avraamidou2016intersections, phipps2010research}. As expressed in the NRC report Learning science in informal environments, ``An important value of informal environments for learning science is being accessible to al'' (National Research Council, 2009). Due to the carefree nature and participant-lead aspect of informal science education, it has often been seen as a key element to spark interest, support student learning, discover the joy and relevance of science in everyday life, and identify science as a possible career path \cite{adams2017informal, anderson2007predators, aroca2011ensino, eratuuli1990experiences, nielsen2009metacognitive, tal2014learning}. 

Besides informal science education, a number of other terms are used to describe activities whereby scientists interact with the public, including non-formal education, public engagement, and outreach. While these are often used interchangeably, it is important to define and describe the differences between these designations -- with the recognition that variations routinely occur with different authors and regions. Non-formal education activities are those that have an established framework but take place outside formal learning structures ({\it e.g.}, classrooms). For example, training activities in the arts or in sports often follow an established syllabus but have few (if any) formal assessments, as they are usually pursued by individuals striving to improve a particular set of skills as a hobby rather than as a career. Courses for senior citizens are another popular example of non-formal educational opportunities. 

Informal education also happens outside of formal learning structures and learners also choose to participate. However, there is not a particular objective or framework defined. Informal learning happens constantly, as it is based on the individual's experience and in some occasions the learning is happening without the individual explicitly choosing it, like reading an article or listening to a radio report. In some cases can be more deliberate, like visiting a museum, attending a summer camp, or downloading a science podcast. While a broad set of activities are classified as informal education, ``outreach'' is a more narrow term that was previously used to describe the interactions of scientists and the public. Unfortunately, the connotation of this word implies a one-way interaction wherein a knowledgeable scientist provides information to the recipient audience who may otherwise not know, be aware of, or have access to this information. This one-way interaction between facilitator and audience, referred to as a deficit model, is however no longer favored: there has been a recent shift in the community to reframe this model and design activities wherein facilitators and audience are mutually engaged, exchanging information and knowledge. Referred to as the two-way interaction model, this reframing has led to more dynamic events, which are now referred to as ``public engagement'' rather than by the term ``outreach''. 

In our discussion of informal science education activities, we will follow this trend by emphasizing activities where there is a mutual interaction between the different stakeholders. {\bf For brevity, we will also use the acronym ISEAs for ``informal science education activities''}. 

\section{Value of Informal Education for Stakeholders}

There are four stakeholders that benefit from ISEAs -- the audiences, the researchers, the institutions, and the field of physics. In support of our effort to have ISEAs considered for career advancement decisions, we will describe the value of ISEAs to each of these stakeholders in this section.

\subsection{The Audiences}

The benefits of classroom science education to audiences have been well documented. These include, for example, enhanced critical thinking \cite{ainsworth2010formal}, cultivating a passion for learning, cross-discipline advances, preparation for the future, and economic benefits ({\it e.g.}, career opportunities) \cite{Connections2021, eng201110};  as well as health benefits ({\it e.g.}, vaccines), understanding public policy \cite{marincola2006public}, and overall science literacy \cite{archer2015science}. 

Not only do ISEAs supplement all these benefits of classroom science, they do so without the constraints of curriculum, text books, tests, metrics, or oversight by school boards or other organizations \cite{falk2005using, avraamidou2016intersections, NAP12190}. This freedom has led, in some cases, to extremely creative and meaningful learning experiences \cite{doering2015fostering, oner2016stem}. When two-way interaction methods are employed, the resulting dialogues can bring science ``alive'', spark interest, and stimulate curiosity in a way that is hard to replicate in classroom curricula designed to cover material on standardized tests and end-of-year exams \cite{avraamidou2016intersections, NAP12190}. For these reasons, ISEAs can inspire students to vigorously pursue classroom science studies. Further, for discussions on critical public policy issues \cite{Fleming2018} such as climate change, nuclear energy, or vaccine efficacy and safety, direct person-to-person engagement between a researcher and the public in an open and accessible forum can change hearts and minds in ways not possible with a journal article or conference lecture that will not be seen by the general public. As scientific advances permeate our daily lives more each day, the audiences benefit from having more access to science in the manner only possible through ISEAs.

\subsection{The Researchers}

Many physicists, and some of their supervisors, view the time spent on ISEAs as time ``lost'' from productive research \cite{andrews2005scientists, martinez2016has}. However, there are numerous ways in which researchers can benefit from their involvement in ISEAs, including: improving their communication skills \cite{hinko2013impacting, illingworth2015developing}; advancing their research; expanding their abilities to mentor, teach and train the next generation of scientists \cite{adams2017informal, avraamidou2016intersections, Bennett2017, hinko2016characterizing}; identifying diverse career paths \cite{laursen2012impact}; and satisfying stakeholder ``societal impact'' requirements \cite{NSF2021}. The origins of these benefits lie in the preparation for, as well as in the participation in, ISEAs. With regard to preparations for engagement activities, researchers must take time away from their regular research duties to first reflect on all aspects of their work -- including their overarching motivation, short-term goals, methodologies, results, impact, and next steps -- and then to formulate clear and concise descriptions of each. With regard to participation, researchers must utilize these formulations to engage in two-way interactions with audiences. We will examine in detail the benefits that this reflective thinking, message formulation, and audience engagement have in their communication, research, teaching, and societal skills. 

\subsubsection{Improving Communication}

Since informal science education activities (ISEAs) are based on communication, researchers who are active in ISEAs can expect to hone their communication skills \cite{gagnon2017addressing, hinko2013impacting, illingworth2015developing}. This includes message crafting (creating, revising, streamlining, and rehearsing),  conceiving analogies and examples to better describe your work, adapting messages to audiences with different backgrounds and interests, responding to unexpected questions, and addressing the critical question ``Why should anyone care about your research?'' These skills, especially the ability to place detailed research work into a much broader context and show its relevance to everyday life, obviously benefit researcher communications with funding agencies, academic departments, review panels, technical audiences, colleagues, and other stakeholders, and can also be invaluable for writing journal articles. 

\subsubsection{Advancing Research}

Beyond improving communication skills, research efforts can be advanced by participation in ISEAs -- specifically, in research direction, methodologies, and a researcher's overall knowledge of the field. For example, the reflective thinking needed to prepare for ISEAs can lead to changing research directions in a way that differs from traditional mechanisms ({\it e.g.}, discussions with collaborators and colleagues, examining the work of competitors, exploring related fields, lack of resources, etc.) 

Another benefit of ISEAs is that engagement with non-experts can lead to ideas that differ from standard methodologies in your field. Such discussions can lead to unexpected inspirations, through ``outside the box'' thinking that can alter both detailed approaches as well as overall research directions. Audience interaction can also lead to new perspectives on priorities, assumptions, and conclusions that may not otherwise occur. This may be particularly apparent in applied physics areas, such as certain aspects of materials science and biomedical physics, where research directions may be driven by pragmatic societal needs and goals rather than pure curiosity.

Finally, because physicists are often asked to engage with the public on topics outside of their primary research area, ISEA preparation and participation can be a gateway to expanding a researcher's physics understanding. It can also enable the discovery of previously unrecognized connections between their research and other areas, connections that could advance their research in unexpected and exciting ways.

There have been a number of studies that have quantified the impact of ISEAs on research productivity. In a study of researchers at a sustainability research center, \cite{kassab2019does} found statistically significant and moderately positive correlations between four out of six types of public engagement activities and the number of research publications. This agrees with other studies finding a positive effect of ISEAs on research performance \cite{bentley2011academic, jensen2008scientists, van2012bench, van2011entrepreneurial}, and is consistent with other studies that have found no negative impact on research \cite{gulbrandsen2005industry, mostert2010societal}. 

\subsubsection{Improving Mentoring, Teaching, and Training}

Because communication skills are absolutely essential to effectively mentor, teach, and train the next generation of scientists, the improved communication skills resulting from two-way engagement-based ISEAs will surely benefit these training efforts. The reflective thinking needed to craft presentations for ISEAs will aid in crafting succinct messages that will make training much more potent, and will greatly assist researchers to impart the broader context and overall motivation for their work to younger scientists. This last point is particularly important, as young scientists who are often overwhelmed with learning the daily research tasks may lose their incentive without knowing the impetus for those tasks. More broadly, ISEAs train scientists to explain their work effectively and with enthusiasm to those unfamiliar with the scientist's field of research - an obvious asset for teaching students in more formal education settings, particularly at the undergraduate and junior graduate levels. Furthermore, ISEAs play an important role in launching the careers of researchers. A recent study \cite{rethman2021impact} used surveys and interviews of over 100 undergraduate students and found that students who facilitated informal physics programs reported positive benefits to their communication, teamwork and networking, and design skills. This helped develop their physics identity, their feeling of belonging to the physics community, and their career skills. Another study \cite{Prefontaine2020} identified interactions with the audience, working in teams, thorough training, clear mission, and supportive schedule as key structures in ISEAs that help build the physics identity of undergraduate participants. 

\subsubsection{Satisfying Societal Impact Requirements}

Funding opportunity announcements from certain agencies ({\it e.g.}, the U.S. National Science Foundation \cite{NSF2021}) require that responding proposals include a portion directed at societal impact. While the training of graduate students was, in past years, often sufficient for such broader impact components, more elaborate efforts to demonstrate how the proposed research serves the needs of society are now prevailing. In many cases, establishing ISEAs tied to the proposed research topic can satisfy, or serve as a critical component of, the societal impact requirements of a proposal. Furthermore, researchers who regularly participate in ISEAs will likely have a propensity for creating new activities that may significantly improve the societal impact component of their next proposal. 
 
\subsection{The Institutions}

Institutional support for informal science education activities (ISEAs) is certainly critical for their success. This support includes, but is not limited to, facilitating, publicizing, hosting, sponsoring, and funding these activities, as well as encouraging their employees to participate. Research examining the landscape of informal physics programs in higher educational institutions (HEI) across the US have, however, found that the vast majority of these activities are not run by tenure system faculty, but rather by staff, students, and instructors. As these facilitators typically have substantially less clout in the hierarchy of academic institutions, ISEAs often exist on the fringe of awareness and resources within the physics culture. 

To help reverse this trend, we note that there are many benefits of ISEAs to the host institution. For example, successful ISEAs can be invaluable recruiting tools -- increasing the number of junior researchers and (at academic institutions) the number of students studying science. As an example of the latter, a recent study of North Carolina State University students found that participants in the North Carolina Science Olympiad, an activity for North Carolina high school students, were significantly influenced in their choice of colleges, their choice of majors, and their decision to go into a STEM field \cite{NCSU2018}. 

Institutions that run successful ISEAs can also significantly enhance their role as ``good neighbors'' in their community. One example of this is the School for Science and Math at Vanderbilt University, which provides a research-based elective STEM curriculum for high school students in Nashville, Tennessee. The program has elevated achievement scores of participants, has produced 27 Intel and Siemens semifinalists and regional finalists, and has resulted in conference presentations and scientific publications for some participants \cite{eeds2014school}. 

ISEAs can not only give institutions increased visibility in their local communities, they can also raise their reputation in the research world. This can come, for example, from enhanced research funding: the careful crafting of messages by facilitators of ISEAs can form the foundation of successful research proposals. Also, as discussed above in Section 3.B.1, numerous studies have shown positive correlations between ISEA participation and research productivity. Finally, ISEAs often satisfy the ``societal benefits'' component that is required in many proposals ({\it e.g.}, NSF \cite{NSF2021}). 

\subsection{The Field of Physics}

There are numerous benefits of ISEAs to the field of physics, and of science in general, including growing the workforce, growing the general public support for research results and methodology, and changing the direction of research and development. We provide details of each of these aspects below. 

\subsubsection{Recruiting and Equity/Diversity/Inclusion}

Recruiting the next generation of physicists, critical for the health of our field, requires substantial effort; it is not a passive endeavor. For example, researchers who simply wait for potential summer interns to contact them are at a significant disadvantage to those who take action to bring in new members to their team. ISEAs have been shown, through a large body of research \cite{barton2013science, maltese2010eyeballs, rahm2016case, zimmerman2012participating, carlone2007understanding, godwin2016identity, varelas2015explorations, archer2010doing, Hazari2019}, to be an excellent mechanism for recruiting. Specifically, these works show that participation in ISEAs are correlated with increased interest in science, development of a science identity, and career intentions. Especially because they can be creatively designed and have no tests or metrics, ISEAs are ideally suited for ``sparking the interest of future generations of scientists'' in a non-threatening and enjoyable manner. 

Furthermore, there is research within the Physics Education Research (PER) field that have looked at the positive impact that facilitating ISEAs has on the students who organize them, including fostering physics identity, development of communication and pedagogical skills, sense of belonging, and confidence among peers \cite{mullen2019should, Prefontaine2020, fracchiolla2020participation, fracchiolla2020community, rethman2021impact, hinko2016characterizing, bennett2020refining, adams2017informal, Hazari2019, king2019black}.

These works have prompted a steady increase in the number of informal physics education programs, resources, and public campaigns, mostly targeted at youth from underrepresented minorities. This focus is warranted because disparity of representation in physics has been an issue for decades. Increasing the number and diversity of students who choose physics as a career has become a key objective nationwide -- to guarantee that the membership of the physics community reflects the rich diversity of the communities it serves, and can benefit from a heterogeneity of thought processes and experiences. 

Because this recent focus on increasing equity, diversity, and inclusion (EDI) holds for all fields, and the pool of applicants is finite, it is more important than ever to design ISEAs that attract underrepresented groups to physics. While the number of students from underrepresented groups that enrolled in physics has increased, in part due to engagement efforts, the trends for students who graduate and pursue a career in physics have unfortunately not changed significantly \cite{Funk2018, Mulvey2020, riegle2019does}. Prior research \cite{godwin2013development, hazari2010connecting, hyater2018critical, hyater2019deconstructing, hazari2020context} demonstrates that developing a physics identity is a key factor in attracting and retaining students in physics, and participation in informal physics spaces can support and foster the development of a physics identity for both audience and facilitators \cite{NAP12190, fracchiolla2020participation, Prefontaine2020, adams2017informal}. 

It should be noted, however, that many who design and facilitate these programs may possess academic experience but lack resources or research-based knowledge on best practices designing, implementing, and assessing the programs, particularly from a cultural competency perspective, putting at risk their efficacy to achieve diversity within physics. This suggests that improving EDI in the field of physics would be well served by researchers working collaboratively with physics education experts in designing ISEAs targeting underrepresented groups. 

\subsubsection{General Public Support for Science}

Science and the technology it enables are major driving forces behind every successful nation, engender healthier, longer, richer lives in the societies that embrace them, and reveal and celebrate the mysteries of the world around us. These successes are reflected in public opinion: 73\% of Americans believe science has a mostly positive impact on society \cite{Funk2020}. Because much scientific research is publicly funded, maintaining and improving that positive public support will be critical to sustain and grow both future research efforts and scientific manpower. 

To this end, a body of peer-reviewed scientific literature on the value of public engagement around science as a means for boosting public acceptance of science -- and importantly, how to do it successfully -- is emerging. A number of different models for ISEAs have been identified \cite{fracchiolla2019characterizing}. \cite{archer2015science} introduce the notion of science capital and suggest framing science engagement in terms of maximizing that capital. \cite{dawson2014not} stresses that integrating low-income ethnic minority groups into public engagement on science requires looking beyond eliminating certain specific barriers, such as cost, to additionally incorporating cultural and linguistic inclusivity. Especially crucial for successfully engaging adult audiences, particularly around contentious science issues, is overcoming the deficit model of science outreach, which is based on a belief that the public is ignorant and seeks to remedy that flaw by providing more and better information; such an approach is inherently paternalistic and reflects a narrow conceptualization of knowledge ({\it e.g.}, \cite{phillips2013really}) that is likely to alienate or offend audiences. 

Beyond research on informal education, some quantitative studies of social interactions also provide important clues on how to develop effective strategies for public engagement on difficult science topics. For instance, \cite{williams2015network} applied complex network theoretic analysis to metadata from social-media discussions of climate change science. They found that the most extreme views were reinforced by homogeneous echo chambers largely devoid of argument, whereas more moderate views evolved in social-media interactions that breached cultural silos, leading to both increased debate but ultimately also decreased polarization. Studies like those described above collectively help provide a framework for formalizing and scaling up ISEAs in a way that can increase public approval of science and scientists. 

An expansion of such efforts is especially needed in light of new challenges that threaten public support for science from several directions. High-profile examples of such challenges include: fears of genetic research and related objections to GMO foods and mRNA vaccines; policy solutions for climate change that are often at the expense of workers and the poor, and similarly, social injustice in some science-based movements like environmentalism (see, {\it e.g.}, \cite{Gross2018, shammin2009impact, LaChance2018, Mock2019, Fears2020}; the recent worldwide rise of populist movements, which are by definition skeptical of experts -- including scientists; and skepticism of the public-health value of masks during the COVID-19 pandemic, which arose from a variety of sources ranging from partisan political drivers to inept science communication that conflated rigorous medical advice with public health management goals \cite{Tufekci2020}. Low-income communities and people of color can be disproportionately affected by these challenges, such as the deeply immoral and racist Tuskegee vaccine experiments that killed over 100 African-Americans in the mid-20th century, eroding the trust these communities have in scientists and resulting in their having substantially lower approval ratings for science \cite{Funk2020}. 

This worsening distrust of scientists is an obvious threat to the scientific enterprise and the societal benefits it can provide. A crucial way to address this unfortunate trend is to enlist more scientists to participate in ISEAs, so that they may effectively serve as ambassadors for science. However, it will be essential to train these new ambassadors so that their efforts improve, rather than erode, public trust in science. For example, training to emphasize engagement over one-way interactions, to focus on the science over public policy or partisan political statements, and to be sensitive to the metaphysical belief systems of the public nationally and globally, perhaps especially indigenous peoples \cite{Fleming2018}, will significantly elevate the quality and positive impact of enhanced engagement efforts.

\subsubsection{Direction of R\&D}

As discussed in Section 3.B.2 above, preparation for and participation in ISEAs may result in changes of the direction of research by individual physicists based on their reflective thinking or on suggestions by (or inspiration from) non-experts. Such a change in direction for a single research effort could, in principle, have a profound and long-lasting impact on a physics subfield. Again, this might be particularly relevant for certain applied physics areas, where research directions may be primarily driven by pragmatic societal needs and goals. Direct engagement with the public and with professional experts in fields other than physics can, additionally, be critically important to identifying and prioritizing knowledge gaps and requirements.

\section{Examples of Successful Engagement Programs}

To illustrate some of the concepts discussed above, we describe a number of examples of ISEAs in different physics subfields that illustrate the two-way engagement model interacting with the public. These include CERN (particle physics), Michigan State University (nuclear physics), Texas A\&M University (modern physics), University of Michigan (general physics), the University of Illinois at Urbana-Champaign (general science), Rutgers University (general physics), and the Princeton Plasma Physics Laboratory (general science), and Science Cafes (general science). 

CERN has an extensive Education, Communication, and Outreach effort \cite{CERN2021} that includes websites, social media, media visits, press releases, programs and activities for teachers and students, guided tours, exhibitions, the Arts AT CERN programs, and more that include photos, videos, and animations. Additionally, a new flagship education and outreach center, the CERN Science Gateway, is under construction and will open in 2023. It will feature immersive hands-on exhibitions, education laboratories, and events for international audiences of all ages. 

Michigan State University has established a set of ISEAs in nuclear physics that reach a broad age range and cover a wide range of activities \cite{Spyrou2021}. These include GUPPY (grades 4-6), MST@MSU (grades 7-10), PAN@MSU (high school), PAN for high school teachers, a High School Honors Science Program, NS3 -- Nuclear Science Summer School (college), Research Assistantships (college), Summer Research Experiences/REU (college), Conference Experiences for Undergraduates (college), and a variety of workshops, schools, conferences, and career opportunities for graduate students. They also offer laboratory tours (4000 visitors/year), open houses (~3000 visitors), public talks, school visits, science festivals. They have also established an infrastructure overseeing these activities, including coordinators, faculty interface, committees, web designers, and more. 

Texas A\&M University \cite{TAMU2020} has 14-year running Saturday Morning Physics program, where Texas high school students learn about modern physics research topics including cosmology, relativity, dark matter and neutrinos, nuclear physics in stars and in medicine, cold atoms and nanodevices. The students receive certificates for sustained attendance, and high school teachers are especially invited to participate. 

Since 1995, the University of Michigan has run a Saturday Morning Physics program \cite{UMICH2021} designed for general audiences. It features faculty members describing their research in non-technical terms, employing multimedia presentations with hands-on demonstrations, slides, videos, and computer simulations.

The Department of Physics at the University of Illinois at Urbana-Champaign has run a traveling science show called the ``Physics Van'' since 1994 \cite{ILL2021}. The van visits schools in many states and performs programs to excite school kids of all different age levels in science. These include assembly-style, classroom workshops, mobile exhibits, and more. They are invited by colleges/universities, research institutions, museums, and private organizations to put on their programs. 

The Rutgers Faraday Holiday Children's Lecture series has been running for over 20 years \cite{RUTG2020}. These holiday physics shows designed for the public have entertained approximately 25,000 children and adults. They are energetic, sometimes explosive hands-on demonstrations that cover matter and motion, Newton's laws, and other topics found in a typical physics semester course. 

The Princeton Plasma Physics Laboratory \cite{PPPL} has been running the Science on Saturday public lecture series since 1991. It features high-school level talks given by experts in diverse fields of science. The audience varies in ages and academic backgrounds and some participants have been joining since their early years. Another event hosted by PPPL is the Young Women's Conference (YWC). The YWC brings together Middle and High School girls for a one day event featuring booths and speakers from different STEM fields and institutions. The YWC quickly outgrew the PPPL site and is hosted yearly in the Princeton University campus where it gathers ~1000 participants a year.

Science Cafes \cite{SCICAFE2021} are a series of informal gatherings for science discussions held in bars, restaurants, and other casual settings. The events feature a moderator and a scientist and the showing of a short video designed to initiate a lively conversation with the audience. With a goal of igniting a passion for science among the attendees, it is common for the audience to take over and be fully engaged, and the topic to meander towards their mutual interests. The wide variety of Science Cafe venues, topics, audiences, and online resources have resulted in tremendous growth of this grassroots movement -- with hundreds of these series now established across the USA and internationally. 

\section{Evaluating Informal Education Efforts for Career Advancement}

There are many types of ISEAs, ranging from public lectures and events to laboratory tours and school visits, from workshops and summer schools to podcasts and videos, from books and media interviews to comic strips and physics kits and more. Some of these activities reach thousands of participants, and others just a few, but all have the potential to bring about profound changes in lives, institutions, and our field. The level of effort required to carry out such a diverse collection of activities clearly varies widely in initiative, commitment, and time. For this reason, it is not practical to establish metrics that could be used to evaluate these efforts for hiring or career advancement. This same lack of metrics also holds for the service work ({\it e.g.}, chairing a department, serving on or leading a committee, conducting peer reviews, organizing conferences, authoring studies, arranging colloquia, and more) that is required for career advancement at most academic institutions. Instead of metrics, evaluations of service work are often made by letters of recommendation that detail the initiative, commitment, time, and impact of these activities. We recommend a similar modality for evaluating ISEAs for career advancement. A ``statement of Informal Education Efforts'' written by candidates can be part of their career advancement portfolio, and letter writers can be instructed to comment, when appropriate, on such efforts. In particular, in lieu of metrics, we provide a representative list of aspects of ISEAs that could be used for career advancement evaluations. These aspects include Initiative, Creativity, Audience size, Duration, Interactivity, Impact, Funding, Publicity, and Longevity. We briefly discuss each of these from the perspective of an evaluator making the case for hiring or career advancement. 

{\it Initiative} -- Conceiving and executing an ISEA often requires significant time and effort, but can result in an event that best utilizes the strengths and background of the presenter to be most impactful.

{\it Creativity} -- Novel concepts in ISEAs have the potential to inspire new ways of learning and possibly launch new informal education paradigms. They can also demonstrate what does not work, and hence can be valuable learning opportunities for facilitators. 

{\it Audience} -- In some cases, the impact of an event scales with the number of participants. Larger events often require an attention to content and logistical details to maximize the potential to reach all audience members. Alternatively, the intimacy of few-to-one or one-to-one interactions can provide tremendous opportunities for learning and mutual engagement. Furthermore, the composition of the audience members should be considered. For example, engaging with a small number of high school teachers may have the same impact as directly presenting to a much larger number of students.

{\it Duration} -- Some activities may require extensive preparation to condense complex science concepts into a short time frame, while others may require covering multiple topics to keep an audience engaged for longer periods. 

{\it Longevity} --  Establishing a web presence, distributing presentation files, creating take-away items, and collecting audience contact information can all enable an ISEA to have an impact that lasts substantially longer than the original event. 

{\it Interactivity} -- The recent shift towards two-way interaction can not only produce events that energize the audience, they can provide learning opportunities for the facilitators. 

{\it Impact} -- Some methods of determining the impact of ISEAs include audience questionnaires, oral or written feedback from participants, the number of questions asked during and after the event, and the number of new students or young researchers resulting from recruiting events. 

{\it Funding} -- Some ISEAs require facilitators to obtain funding, adding another facet to the commitment and effort to realize the event. 

{\it Publicity} -- Articles about successful ISEAs can bring positive attention to the host institution, can raise the reputation of departments or divisions, and can inspire others to initiate events. Some articles are effectively an additional ISEA that further broadens the audience and impact of the original event. 

Consideration of these aspects of ISEAs may facilitate hiring and career advancement decisions by the host institutions of the facilitators. We note that some of these same criteria are used by APS committees in their decisions to fund ISEAs ({\it e.g.}, through the Mini-Grant proposal review process run by the APS Committee to Inform the Public \cite{APS2021} ). 

\section{Summary}

We have enumerated the many benefits of informal science education activities to the audiences, to researchers, to their institutions, and to the field of physics. These include: enhancing the critical thinking, understanding of public policy, career opportunities, and science literacy of the public; improved communication skills, methodology, and mentoring by researchers; increased science enrollment, visibility, reputation, and  research funding for institutions; and expanded and more diverse recruiting and more public support for the field of physics. Given these benefits, we advocate for the expansion of informal science education activities, especially those that involve two-way engagement with the audience. We also advocate that institutions consider involvement in these activities as part of their decision-making process on hiring and career advancement, and we discuss numerous aspects of these activities that could be helpful in evaluations required for career advancement. 

\Urlmuskip=0mu plus 2mu\relax
\bibliographystyle{unsrtnat}

\bibliography{Engagement_V2}{}

\end{document}